\documentclass[journal=cgdefu,manuscript=article,layout=twocolumn]{achemso}

\usepackage[sc]{mathpazo}
\usepackage{helvet}
\usepackage[utf8]{inputenc} 
\usepackage{graphicx}
\usepackage{xspace}

\newcommand{\grad}{$^{\circ}$\xspace}
\newcommand{\celsius}{$^{\circ}$C\xspace}

\author{Oliver Brandt}
\email{brandt@pdi-berlin.de} 
\author{Sergio Fernández-Garrido}
\author{Johannes K. Zettler}
\author{Esperanza Luna}
\author{Uwe Jahn}
\author{Caroline Chèze}
\author{Vladimir M. Kaganer}
\affiliation[Paul-Drude-Institut für Festkörperelektronik]{Paul-Drude-Institut für Festkörperelektronik, Hausvogteiplatz 5--7, 10117 Berlin, Germany}

\title[Coalescence degree of spontaneously formed GaN nanowires]
{Statistical analysis of the shape of one-dimensional nanostructures: determining the coalescence degree of spontaneously formed GaN nanowires}

\begin{document}

\begin{abstract}
Single GaN nanowires formed spontaneously on a given substrate represent nanoscopic single crystals free of any extended defects. However, due to the high area density of thus formed GaN nanowire ensembles, individual nanowires coalesce with others in their immediate vicinity. This coalescence process may introduce strain and structural defects, foiling the idea of defect-free material due to the nanowire geometry. To investigate the consequences of this process, a quantitative measure of the coalescence of nanowire ensembles is required. We derive objective criteria to determine the coalescence degree of GaN nanowire ensembles. These criteria are based on the area-perimeter relationship of the cross-sectional shapes observed, and in particular on their circularity. Employing these criteria, we distinguish single nanowires from coalesced aggregates in an ensemble, determine the diameter distribution of both, and finally analyze the coalescence degree of nanowire ensembles with increasing fill factor.
\end{abstract}

\subsection*{Keywords}
{\small Keywords: nanowires, nanoparticles, coalescence, statistical analysis}
%\mciteErrorOnUnknownfalse

\section{Introduction}
The compound semiconductors GaN and ZnO tend to spontaneously form dense ensembles of nanowires on a variety of substrates.\cite{Geelhaar_ieee_2011,Schmidt-mende_mattoday_2007} As a direct consequence of the quasi one-dimensional nature of these nanowires, dislocations forming at the interface to the substrate do not propagate along the nanowire axis, but remain at the interface\cite{Tomioka_nl_2008} or bend towards the free sidewall surfaces.\cite{Urban_njp_2013} Single nanowires are thus indeed free of threading dislocations in contrast to heteroepitaxial GaN or ZnO layers.\cite{Trampert_iop_2003,Perillat_iop_2010} Consequently, these nanowire ensembles are attracting much interest for applications requiring single crystals of high structural perfection, such as demanded for light emitting or light harvesting devices.\cite{Li_jap_2012,Howell_nl_2013}   

The nanowire ensembles, however, are of such high density (10$^9 \dots 10^{10}$~cm$^{-2}$) that nanowires close to each other may inadvertently coalesce due to radial growth or due to their mutual misorientation. GaN nanowires undergo radial growth immediately after their formation,\cite{Garrido_nl_2013} and nanowire coalescence may be significant at this stage if their density is sufficiently high. GaN nanowires furthermore exhibit an out-of-plane orientational spread on the order of one degree,\cite{Jenichen_nt_2011,Wierzbicka_nt_2013} promoting additional nanowire coalescence at later stages of growth.

This mutual misorientation also causes the adverse effects of nanowire coalescence. Upon coalescence, a small tilt may be accommodated elastically, but a larger tilt as well as a twist will result in dislocated tilt/twist boundaries reminiscent of small-angle grain boundaries in polycrystalline thin films.\cite{Consonni_apl_2009,Grossklaus_jcg_2013} Coalescence is thus likely to introduce inhomogeneous strain\cite{Jenichen_nt_2011,Kaganer_prb_2012} as well as nonradiative recombination,\cite{Sanchez-Garcia_jcg_1998,Park_nt_2006,Consonni_apl_2009} phenomena one associates with heteroepitaxial thin films rather than with nanowire ensembles. The detrimental impact of coalescence on the structural perfection of dense nanowire ensembles may prevent these nanostructures to realize their full potential for applications.

Despite the fact that nanowire coalescence and its significance have been recognized from the very beginning of the history of GaN nanowire growth,\cite{Sanchez-Garcia_jcg_1998} there have been comparatively few studies investigating this effect in more detail,\cite{Park_nt_2006,Calarco_nl_2007,Bertness_jcg_2008,Garrido_jap_2009,Limbach_joam_2010,Brubaker_jap_2011,Grossklaus_jcg_2013} and only one in which an attempt was made to quantify the coalescence degree $\rho$ of GaN nanowire ensembles.\cite{Consonni_apl_2011} This latter study focused on a determination of the nucleation density of GaN nanowires, and was thus concerned with identifying the number of nanowires participating in the formation of a coalesced aggregate. For this task, the area of each aggregate was divided by the area assigned to a single nanowire, and the coalescence degree was calculated from the ratio between the number of aggregates and the number of constituent nanowires. However, this procedure and its inherent limitations were not discussed in detail. 

The purpose of the present work is to establish objective criteria which may be used for constructing an algorithmic procedure for distinguishing single nanowires from coalesced aggregates. To this end, we analyze the cross-sectional shape of GaN nanowires using top- and plan-view micrographs obtained by field-emission scanning and transmission electron microscopy, respectively. We show that the area-perimeter relationship of the individual nanowires, and in particular their circularity, may be used for the definition of a coalescence degree which is clear and free from ambiguities. Using this definition, we extract the diameter distribution of single nanowires in an ensemble, and investigate the coalescence degree of standard nanowire ensembles as a function of their fill factor. 

\section*{Experimental}
\label{sec:exp}
{\small
 
\paragraph{Synthesis of GaN nanowire ensembles} The GaN nanowires under investigation were synthesized in molecular beam epitaxy systems equipped with a solid-source effusion cell for Ga and a radio-frequency plasma source for generating active N. All growth took place on Si(111) substrates at temperatures between 780 and 820\celsius, and with varying Ga/N flux ratio. In the course of the present work, we have investigated several dozens of different samples. In general, lower substrate temperatures and higher Ga/N flux ratios were found to favor the radial growth of the nanowires (see Ref.~\citenum{Garrido_nl_2013} for as systematic investigation of the latter phenomenon) and thus to increase the coalescence degree of the ensemble. Samples \textsf{A}, \textsf{B}, and \textsf{C} are representative examples for GaN nanowire ensembles with medium, high, and very high area coverage (equivalent to the fill factor of the ensemble) as well as degree of coalescence, respectively, and are used to motivate and introduce methods for the quantitative determination of the coalescence degree (cf.\ Fig.~\ref{fig1}). Sample \textsf{D} is an example of a GaN nanowire ensemble right after nucleation (for further details, see Refs.~\citenum{Pfuller_nr_2010,Gorgis_prb_2012}). The nanowire density is sufficiently low (10$^8 \dots 10^9$~cm$^{-2}$) to render coalescence unlikely. We hence use this sample to obtain an impression about the cross-sectional shape of single free-standing GaN nanowires (cf.\ Fig.~\ref{fig2}). Finally, we use several samples of a growth series for which the Ga/N ratio was systematically varied to elucidate the dependence of the coalescence degree on coverage (for details on a related sample series, see Ref.~\citenum{Garrido_nl_2013}). 

\paragraph{Scanning and transmission electron microscopy}
The morphology of all of these samples was imaged by field-emission scanning electron microscopy carried out in a \textsc{Hitachi S-4800}. Top-view micrographs  were recorded using an acceleration voltage of 5~kV and a magnification of 20000$\times$, covering an area of 30~$\mu$m$^2$ and thus several hundreds of nanowires for a low degree of coalescence. For samples with a high degree of coalescence close to the percolation threshold, at which all nanowires are coalesced into one giant aggregate spanning the entire sample, up to ten of such micrographs were recorded to obtain a sufficiently large number of nanowires. 

For samples with nanowires with a mean diameter of 50~nm and smaller, we observed that nanowires start to bend during exposure to the electron beam, forming clusters which were similar in appearance as the aggregates formed due to nanowire coalescence during growth. To avoid this electrostatically induced attraction between charged nanowires, we recorded micrographs in a \textsc{Zeiss Ultra-55} scanning electron microscope using ultra-low acceleration voltages between 100 and 200~V. To account for the small diameter of these nanowires, several micrographs were recorded with 40000$\times$ magnification. 

Sample \textsf{D} was additionally investigated by plan-view transmission electron microscopy using a \textsc{JEOL JEM-3010} operating at 300~keV. The plan-view cross-sectional specimens were prepared using mechanical thinning, dimpling and ion-milling using a \textsc{Gatan PIPS}\texttrademark. The details of this preparation were modified with respect to planar structures to account for the peculiar mechanical properties of the GaN nanowires. In particular, the preparation of plan-view specimens suitable for high-resolution imaging requires an initial mechanical stabilization of the nanowires and a limited exposure to the ion-milling process.\cite{Luna_unpubl}

\paragraph{Micrograph analysis}
Both top-view scanning and plan-view transmission electron micrographs were analyzed with the help of the open-source software \textsc{ImageJ}.\cite{Schneider_nmet_2012} We used the default algorithms for the determination of area and perimeter since we found them to be sufficiently robust with respect to the finite resolution  of the experimental micrographs. Prior to thresholding the images as required for this analysis, the micrographs were despeckled and enhanced in contrast.   
}     

\section*{Results and discussion}

\begin{figure*} 
\includegraphics[width=16cm]{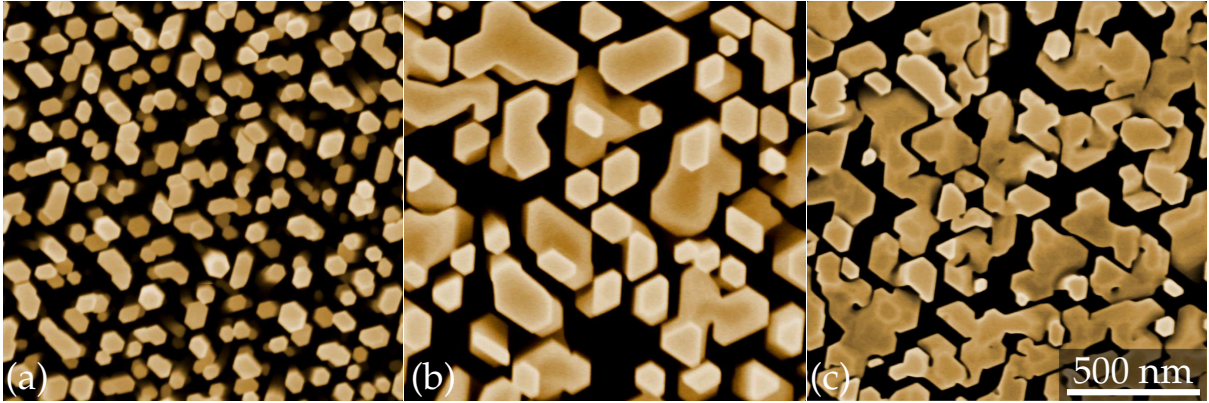} 
\caption{Top-view scanning electron micrographs of the three GaN nanowire ensembles on Si(111) denoted as samples \textsf{A}, \textsf{B}, and \textsf{C} with an area coverage of (a) 0.4, (b) 0.55, and (c) 0.72, respectively. The scale given in (c) applies to all images.   
} 
\label{fig1}
\end{figure*}

Figures~\ref{fig1}(a)--\ref{fig1}(c) show top-view scanning electron micrographs of three GaN nanowire ensembles on Si(111) denoted as samples \textsf{A}, \textsf{B}, and \textsf{C}, respectively. It is immediately evident that the size of the objects as well as the complexity of their shape increases with increasing coverage from sample \textsf{A} to \textsf{C}. Intuitively, one associates these changes with the coalescence of nanowires, and this impression turns out to be correct. However, it is surprisingly difficult to quantify this intuitive interpretation, and it is worth to analyze the implicit assumptions underlying it. 

% FIG

Spontaneously formed GaN nanowires crystallize in the  thermodynamically stable hexagonal modification of GaN. These nanowires are known to grow along the polar GaN$[000\bar{1}]$ 
direction\cite{Hestroffer_prb_2011,Garrido_nl_2012} and to exhibit flat GaN($000\bar{1}$) (\textit{C}-plane) top facets and atomically abrupt non-polar side facets formed by GaN($1\bar{1}00$) (\textit{M}) planes.\cite{Trampert_iop_2003,Bertness_jem_2006,Largeau_nt_2008,Cheze_apl_2010} Density functional theory has shown the \textit{M}-plane to have a lower surface energy compared to the other non-polar plane, namely, GaN ($11\bar{2}0$) or \textit{A}-plane.\cite{Northrup_prb_1996,Lymperakis_prb_2009} Consequently, the \textit{M}-plane is found to emerge in the equilibrium crystal shape of GaN.\cite{Bryant_jcg_2013} Following the same reasoning, a single nanowire should tend to minimize its surface, and its cross-sectional shape should thus approach a perfect regular hexagon. The distinction between single nanowires and coalesced nanowire aggregates would then be straightforward: all shapes deviating from a regular hexagon would originate from a coalescence event.

% FIG
\begin{figure*}[t]
\includegraphics[width=12cm]{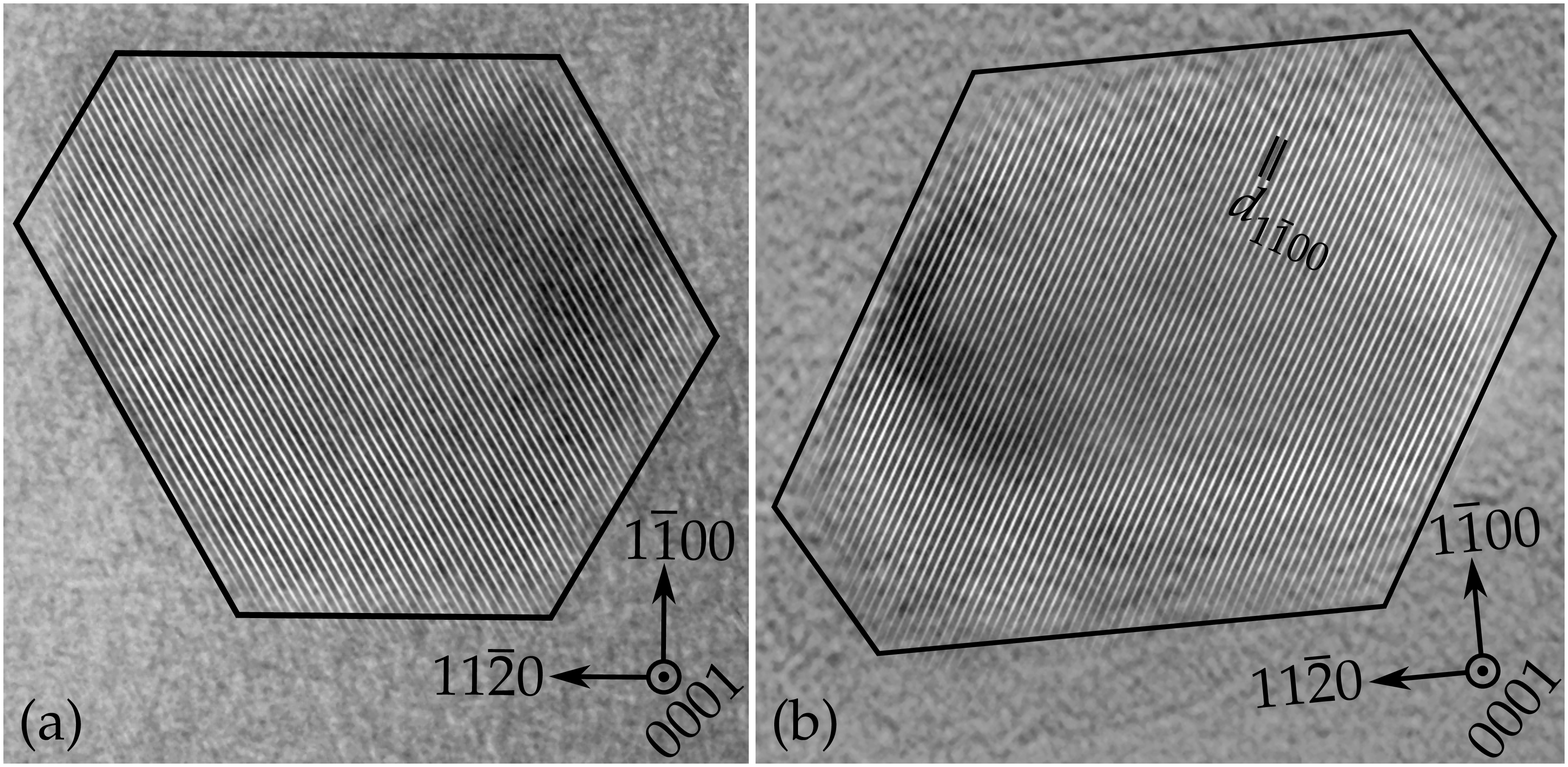} 
\caption{Plan-view high resolution transmission electron micrographs of two GaN nanowires [(a) and (b)] from sample \textsf{D} in their early stage of growth recorded along the [0001] zone axis. The $\{1\bar{1}00\}$ lattice planes with a distance $d_{1\bar{1}00}$ of 0.276~nm are resolved in these micrographs. The corresponding lattice fringes are digitally enhanced by a Fourier filter selecting only the relevant spatial frequency. The irregular hexagon tracing the periphery of the respective nanowire visualizes the well-developed $\{1\bar{1}00\}$ facets delimiting the nanowire. The crystallographic orientation of these facets is shown in the figures.} 
\label{fig2} 
\end{figure*} 

We indeed observe many clear hexagonal shapes in Fig.~\ref{fig1}, and particularly so in Fig.~\ref{fig1}(a). It is this fact which one (correctly) associates with a low degree of coalescence. However, the distinction is much less clear-cut than intuitively believed. A closer examination of Figs.~\ref{fig1}(a) and \ref{fig1}(b) shows that we rarely, if ever, observe perfectly regular hexagonal shapes but instead irregular, distorted hexagons as well as more complex elongated, kinked or branched shapes which dominate the morphology of sample \textsf{C} [cf.\ Fig.~\ref{fig1}(c)]. These latter objects tend to have larger sizes than the more regularly shaped ones, but it is important to note that even the most regular shapes in Fig.~\ref{fig1}(a) exhibit a wide distribution of diameters. 

The crucial question which arises from these experimental findings is whether or not the irregular hexagonal shapes are also the result of coalescence. In order to answer this question, we next examine selected nanowires of sample \textsf{D}. As explained in Section \ref{sec:exp}, this sample exhibits a sufficiently low nanowire density to facilitate the investigation of single nanowires right after their formation.

Figure \ref{fig2} displays plan-view high resolution transmission electron micrographs of two GaN nanowires [(a) and (b)] from sample \textsf{D} in their early stage of growth. The diameter of both of these nanowires amounts to about 15~nm, just slightly above the critical diameter for nanowire formation of 10~nm experimentally established by Consonni \emph{et al.}\cite{Consonni_prb_2011} Nanowire coalescence at this stage of growth is certainly a very unlikely event, and can even be ruled out for the two exemplary nanowires shown in Fig.~\ref{fig2}. As the $\{1\bar{1}00\}$ lattice planes are resolved in the micrographs, the significant mutual misorientation of nanowires originating from independent nuclei would be plainly obvious in these high-resolution micrographs.

The clearly resolved $\{1\bar{1}00\}$ lattice planes also show that both nanowires are delimited by \textit{M}-plane facets which are thus found to be indeed atomically abrupt (see the edges parallel to the $\{1\bar{1}00\}$ lattice planes). Most important, however, is the observation that neither of these two single nanowires forms a regular hexagon, despite the fact that they are evidentially not coalesced. As a matter of fact, we have investigated more than a dozen of such single nanowires, and all of them are characterized by an irregular hexagonal cross-section even at this early stage of growth. For all of these nanowires, we rarely observed atomically sharp corners, but rounded ones as clearly visible for the two nanowires displayed in Fig.~\ref{fig2}. Apparently, the formation of atomically sharp corners is energetically unfavorable.

The irregular cross-sectional shape of single GaN nanowires can be understood within the frame of the unified growth model we have proposed recently.\cite{Garrido_nl_2013} Following the shape transformation of the initial nucleus to the nanowire morphology,\cite{Consonni_prb_2011} the nanowire collects a significant amount of Ga due to the diffusion of Ga adatoms on the substrate in addition to the directly impinging Ga flux. The Ga adatom concentration on the nanowire top facet will then, in general, exceed the available concentration of N adatoms. This excess Ga initiates two-dimensional island formation at the nanowire side facets and subsequent expansion of the island by step flow, and radial growth will set in consequentially.\cite{Garrido_nl_2013}  

In reality, we are not only dealing with one single nanowire, but with an ensemble of them in which each nanowire competes for the Ga adatoms diffusing on the substrate with its immediate neighbors. Since the distance between the individual nuclei is random, the additional Ga flux reaching a single nanowire by diffusion is spatially anisotropic. As a result, the radial growth rate of the six \textit{M}-plane facets differs, and the slowest growing facets will develop the largest edge length. 

This scenario implies that single nanowires may exhibit various irregular hexagonal shapes depending on the spatial location of their next-nearest neighbors. Nanowire coalescence, on the other hand, should result in elongated and branched shapes such as frequently observed in Figs.~\ref{fig1}(a)--\ref{fig1}(c). Several quantities  exist which are sensitive to this change of the compactness of a  shape. One of the most easily obtained and commonly used measures for the compactness is the two-dimensional isoperimetric quotient or circularity $\mathcal{C}$ (for an excellent overview, see Ref.~\citenum{Li_geogr_2013}). The circularity of an object with arbitrary cross-sectional shape is defined by 
\begin{equation}
 \mathcal{C} = \frac{4 \pi A}{P^2}
\label{circularity}
\end{equation}
with the area $A$ and the perimeter $P$ of the cross-section. For regular $n$-gons, this expression simplifies to 
\begin{equation}
 \mathcal{C}_n = \frac{\pi}{n \tan(\pi/n)}.
\label{ngon}
\end{equation}

\begin{figure}[t]
\includegraphics[width=7cm]{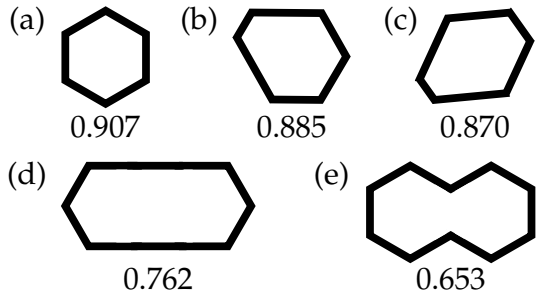} 
\caption{Selected cross-sectional geometrical shapes formed by \textit{M}-plane facets and their circularity. Shapes occuring for single nanowires are shown in (a)--(c). The regular hexagon in (a) exhibits the highest circularity possible for any hexagonal shape. The irregular hexagons in (b) and (c) correspond to those observed in Figs.~\ref{fig2}(a) and \ref{fig2}(b) and are characterized by a slightly lower circularity. The shapes depicted in (d) and (e) are produced by the coalescence of regular hexagons and have the highest symmetry of any conceivable shape produced by coalescence of regular or irregular hexagons. For the former, the coalescence boundary is parallel to GaN($11\bar{2}0$) (the \textit{A}-plane), while it is parallel to  GaN($1\bar{1}00$) (the \textit{M}-plane) for the latter.} 
\label{fig3} 
\end{figure}

Figures \ref{fig3}(a)--\ref{fig3}(c) shows various examples for the cross-sectional shapes of single nanowires. The circularity of the regular hexagon in Fig.~\ref{fig3}(a) as obtained by Eq.~(\ref{ngon}) is the highest possible for any hexagonal shape, but the values obtained for the two cross-sectional shapes observed in Fig.~\ref{fig2} are only slightly lower [cf.\ Figs.~\ref{fig3}(b) and \ref{fig3}(c)] since they are still largely isotropic. In contrast, the circularity of the two shapes shown in Figs.~\ref{fig3}(d) and \ref{fig3}(e) is markedly lower. These two shapes are produced by coalescence of two single nanowires of regular hexagonal cross-section with the coalescence boundary parallel to the \textit{A}- and the \textit{M}-plane, respectively. A vertical offset of the individual nanowires with respect to each other would produce kinked shapes of again significantly lower circularity. 

% FIG

In order to use the circularity as a criterion for distinguishing single nanowires from coalesced aggregates, we need a sensible threshold value $\zeta$ below which all nanowires can be assumed to be coalesced. Figure \ref{fig3} implies that $\zeta_{A} = 0.762$ or $\zeta_{M} = 0.653$ would be such values. It is not clear, however, whether such symmetric shapes actually occur in GaN nanowire ensembles, or if coalescence is not more likely to produce kinked and branched shapes with significantly lower circularity. Furthermore, we have not yet ascertained that the shape of single nanowires in fully developed nanowire ensembles, for which shadowing of the direct Ga flux occurs, will be similar to that observed in the early stage of growth. 

\begin{figure*} 
\includegraphics[width=14cm]{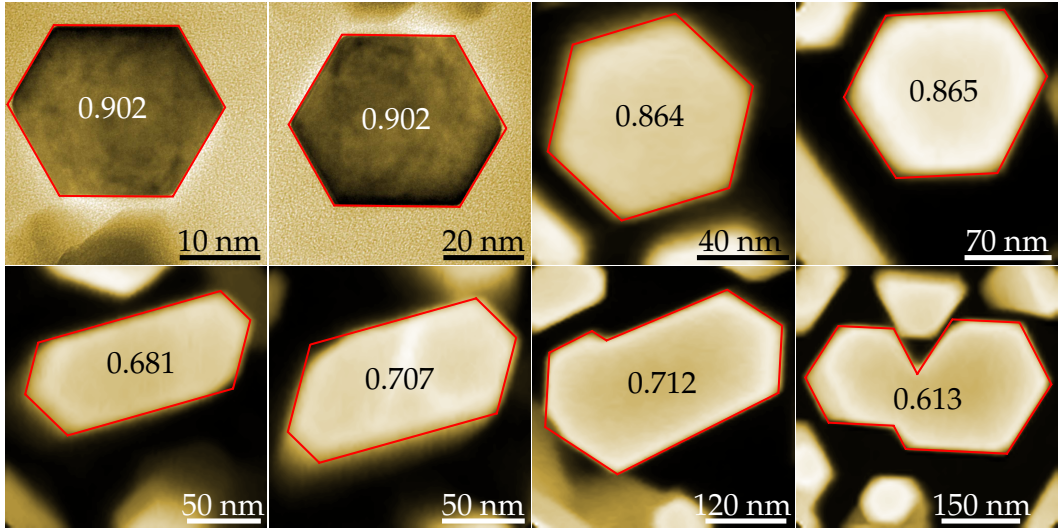} 
\caption{Cross-sectional shapes of GaN nanowires as observed by plan-view transmission and scanning electron microscopy. Each nanowire is labeled by the circularity $\mathcal{C}$ of its shape, which is highlighted by the solid red line surrounding the nanowire. The images have different scales as indicated in the figures. The upper row shows examples of single nanowires from sample \textsf{D} with diameters of 25 and 50~nm ($\mathcal{C}=0.902$), sample \textsf{A} with a diameter of 80~nm ($\mathcal{C}=0.864$), and sample \textsf{B} with a diameter of 140~nm ($\mathcal{C}=0.865$). The lower row shows examples of elongated objects with still comparatively symmetric shape. The left two images are taken from sample \textsf{A}, the right two from sample \textsf{B}.} 
\label{fig4} 
\end{figure*}

We approach these questions by analyzing the shape of nanowires observed in samples \textsf{A}--\textsf{D}. Figure \ref{fig4} displays representative examples of both single nanowires (upper row) and coalesced aggregates of high symmetry and thus high circularity (lower row). Two important conclusions can be drawn from these images. First, cross-sectional shapes close to a regular hexagon are found independent of nanowire diameter in a range of at least 25 to 140~nm. Second, the circularity of coalesced aggregates may exceed 0.7, larger than the value obtained for a rhombus (0.680) or an equilateral triangle (0.605).

% FIG

This latter observation reveals a potential pitfall of the circularity criterion for the detection of coalescence. Specifically, we can construct highly symmetric shapes from a regular hexagon by, for example, shortening the length of two or three of its edges. In the limit of zero edge length, we then approach a rhombus or an equilateral triangle which have lower circularity than some of the shapes produced by coalescence. While we have not observed such extreme cases, we cannot rule out their occurrence, and there is thus a range of circularity values in which both single nanowires and coalesced aggregates may possibly coexist.

We thus suggest an alternative, and partly complementary criterion for the coalescence of nanowires that is also based on their area-perimeter relationship but emphasizes the functional relation between area and perimeter. The area of equilateral shapes, \textrm{i.\,e.}, of regular $n$-gons, is proportional to the square of their perimeter regardless of their circularity. In contrast, the area of linearly extended shapes depends linearly on perimeter, a fact which has been noted and used previously in the study of fractal clusters near the percolation threshold.\cite{Voss_prl_1982} 

Both of the above criteria for distinguishing single nanowires and coalesced nanowire aggregates only require the determination of the cross-sectional area and perimeter of a preferably large number of specimen. For this task, we analyze top-view scanning electron micrographs covering at least several hundreds of spatially separate objects using \textsc{ImageJ}.\cite{Schneider_nmet_2012} 

\begin{figure*}[!t] 
\includegraphics[width=\textwidth]{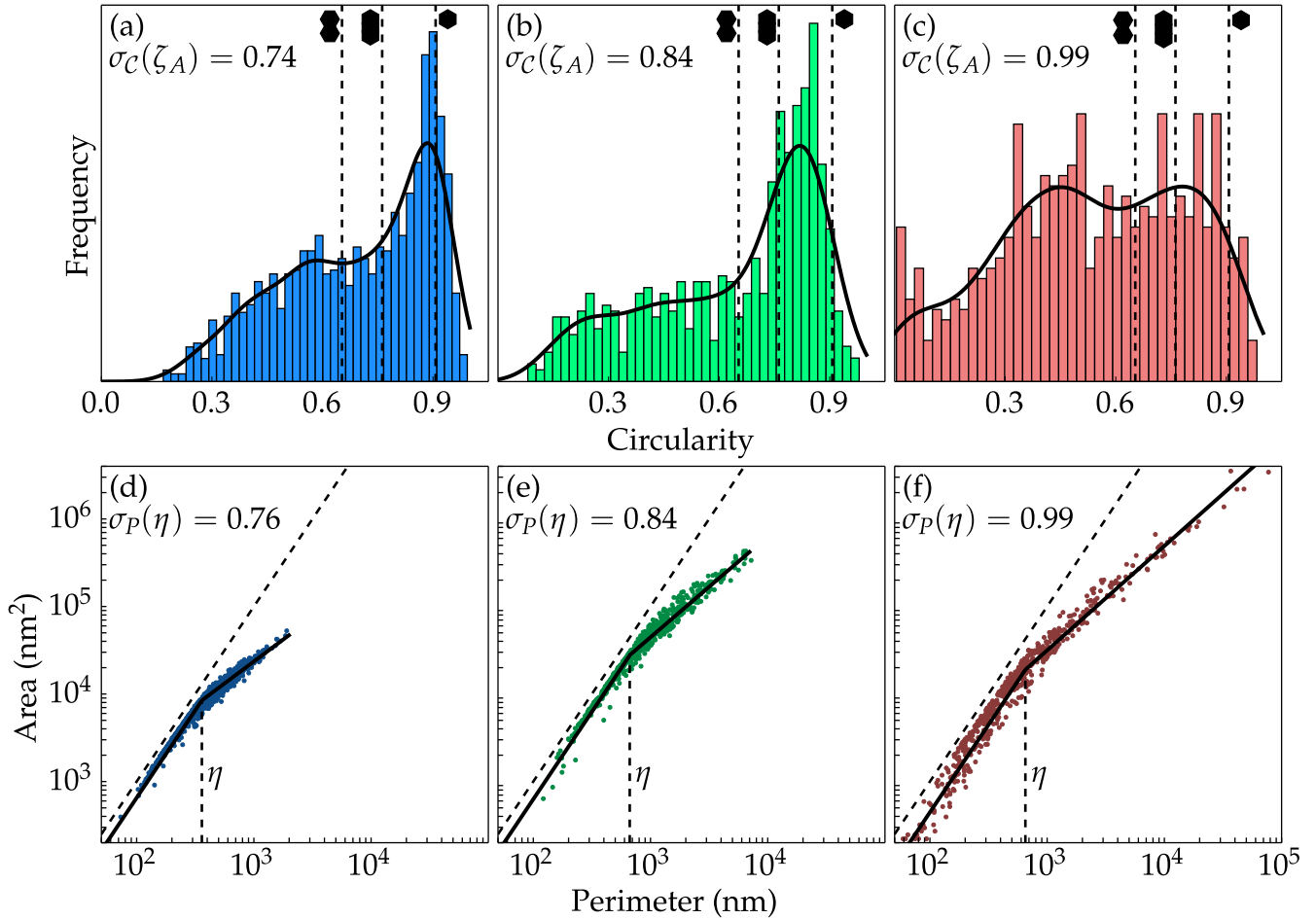} 
\caption{Histogram of circularity and area-perimeter plot for the GaN nanowires from sample \textsf{A} [(a) and (d)], sample \textsf{B} [(b) and (e)], and sample \textsf{C} [(c) and (f)], respectively. The solid line in (a)--(c) shows the kernel density estimation of the respective histogram. The vertical dashed lines indicate the circularity of the geometrical shapes displayed next to them. The solid line in (d)--(f) shows a fit of the data (solid symbols) by Eq.~\ref{ap}, the sloped dashed line a quadratic dependence of the area on perimeter, and the vertical dashed lines indicate the critical perimeter $\eta$. The coalescence degree obtained using either the criterion of a minimum circularity of $\zeta_A=0.762$ [$\sigma_\mathcal{C}(\zeta_A)$] or the criterion of a critical perimeter $\eta$ [$\sigma_{P}(\eta)$] is also provided.} 
\label{fig5} 
\end{figure*}

Figure \ref{fig5} presents the analysis of the data thus obtained for samples \textsf{A}--\textsf{C} based on both the criteria proposed above. The circularity histograms displayed in Figs.~\ref{fig5}(a)--\ref{fig5}(c) are peaked at values corresponding to single nanowires for samples \textsf{A} and \textsf{B}, but are almost uniformly distributed (with a secondary maximum close to zero) for sample \textsf{C}. For all samples, we also observe circularities higher than those expected for a regular hexagon. This effect is primarily caused by the rounded corners of the nanowires (see the corresponding remark in the discussion of Fig.~\ref{fig2}), by their finite tilt, and the finite contrast of the experimental micrographs.

% FIG

We may introduce two different definitions of a coalescence degree. First, we may simply count all objects $O$ with a circularity below a certain threshold $\zeta$, and divide this number $N_{\mathcal{C}<\zeta}$ by the total number of objects $N$
\begin{equation}
\rho_\mathcal{C}(\zeta) = \frac{N_{\mathcal{C}<\zeta}}{N} 
\label{circularity_number}	
\end{equation}  
to arrive at a degree of coalescence $\rho_{\mathcal{C}}$ related to the \emph{number} of nanowires. This definition is useful for the investigation of phenomena and properties for which the number of single nanowires is most relevant, such as, for example, in the study of the nucleation density in  Ref.~\citenum{Consonni_apl_2011}. For samples close to the percolation limit as our sample \textsf{C}, the number of nanowires is limited, and kernel density estimations such as those shown by the solid lines in Figs.~\ref{fig5}(a)--\ref{fig5}(c) will prove useful to smoothen the histogram and to obtain more reliable values.\cite{Rosenblatt_ams_1956,Parzen_ams_1962}

Second, we may sum the area $A$ of all objects with a circularity $\mathcal{C}$ below $\zeta$, and divide this area $\sum_i A = A_{\mathcal{C}<\zeta}$ by the total area $A_T$ of \textit{C}-plane top-facets 
\begin{equation}
\sigma_\mathcal{C}(\zeta) = \frac{A_{\mathcal{C}<\zeta}} {A_T}.	
\label{circularity_area}
\end{equation}  
For an equal length of the nanowires, this latter definition of a coalescence degree $\sigma_{\mathcal{C}}$ corresponds to the degree of coalescence referring to \emph{volume}, and is thus the relevant quantity when examining data from experimental techniques probing the material volume such as x-ray diffractometry and Raman spectroscopy. We will use this latter definition [Eq.~(\ref{circularity_area})] in all what follows. The coalescence degree thus obtained taking $\zeta_A = 0.762$ increases from 0.74 for sample \textsf{A} to essentially 1 for sample \textsf{C} (cf.\ Fig.~\ref{fig5}). When chosing, instead, $\zeta_M = 0.653$ as the divider between single nanowires and coalesced aggregates, $\sigma_\mathcal{C}(\zeta)$ amounts to 0.62, 0.75, and 0.99 for samples \textsf{A}, \textsf{B}, and \textsf{C}, respectively.

The alternative area-perimeter representation of the same data for samples \textsf{A}--\textsf{C} is shown in Figs.~\ref{fig5}(d)--\ref{fig5}(f), respectively. For all samples, the slope of the data changes abruptly from quadratic to essentially linear at a certain critical perimeter $\eta$ (see the dashed lines in Fig.~\ref{fig5}). Counting the number of objects with a perimeter larger than $\eta$, we see that the proportion of nanowires exhibiting a linear area-perimeter relationship increases when going from sample \textsf{A} to \textsf{C}.

Analogously to Eq.~(\ref{circularity_area}), we can define a coalescence degree $\sigma_{P}$ as
\begin{equation}
\sigma_{P}(\eta) = \frac{A_{P>\eta}} {A_T},
\label{equilaterality}
\end{equation}  
where we sum over all objects with a perimeter $P$ larger than $\eta$. To obtain $\eta$, we fit the data with the scaling law
\begin{equation}
A = \alpha\left[ P^2 H\left(\eta - P\right) + \eta ^{2 - \beta} P^\beta H\left(P - \eta\right)\right],
\label{ap}
\end{equation}
where $H$ is the Heaviside function, $\alpha$ a proportionality constant, and $\beta$ is the slope for objects with a perimeter larger than $\eta$. The fit of the data by Eq.~(\ref{ap}) returns values for $\beta$ close to 1 and for $\eta$ as indicated in Fig.~\ref{fig5}. Using Eq.~(\ref{equilaterality}), we then obtain values for the coalescence degree $\sigma_{P}$ very close to those determined for $\sigma_{\mathcal{C}}(\zeta_A)$ given in Eq.~(\ref{circularity_area}) (cf.\ Fig.~\ref{fig5}). As we will see later, the deviation of these values for sample \textsf{A} is a systematic effect: The lower the coalescence degree, the less data points constitute the linear portion of the area-perimeter plot, and the fit will then tend to underestimate the critical perimeter $\eta$.

% FIG
\begin{figure}[t]
\includegraphics[width=\columnwidth]{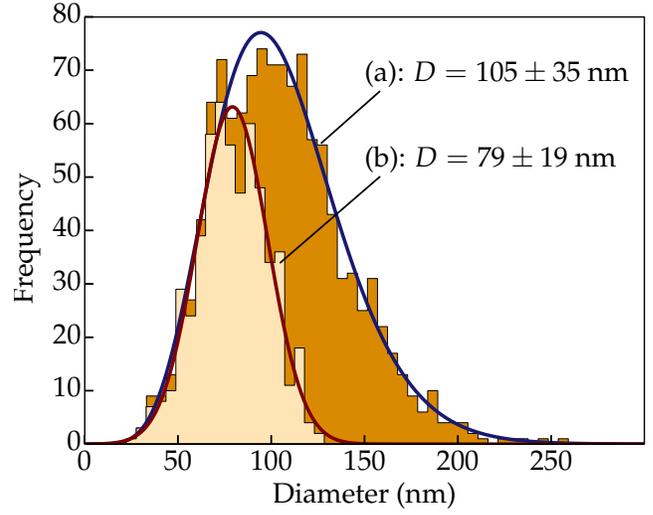} 
\caption{Distribution of the equivalent-disk diameter $D$ of (a) all nanowires of sample \textsf{A} [cf.\ Fig.~\ref{fig1}(a)] and (b) only those with a circularity larger than 0.762. The histograms are fit by (a) a shifted gamma and (b) a normal distribution as shown by the solid lines, yielding the respective mean diameter and its standard deviation indicated in the figure.} 
\label{fig6} 
\end{figure} 

Since we now seem to possess a sensible method for quantitatively determining the coalescence degree of nanowires, we next examine the impact of coalescence on the distribution of the nanowire diameters. Figure \ref{fig6} shows the distribution of the equivalent-disk diameter $D =2\sqrt{A/\pi}$ for sample \textsf{A}, including either (a) all nanowires or (b) only those with a circularity larger than $\zeta_{A}=0.762$. The former is clearly skewed, tailing towards larger diameters, and follows a shifted gamma distribution as observed previously.\cite{Gorgis_prb_2012,Pfuller_apl_2012} In contrast, the latter is symmetric and is represented well by a normal distribution. The asymmetry of the diameter distribution is thus caused by nanowire coalescence, and Fig.~\ref{fig6} is essentially a visual representation of the coalescence degree $\sigma_{\mathcal{C}}(\zeta_{A})$ in that the area outside the normal distribution originates from coalesced aggregates. Note that the diameter variation of the single nanowires is still significant: the full-width-at-half-maximum of the distribution shown in Fig.~\ref{fig6} amounts to 45~nm, \textrm{i.\,e.}, 76\% of the single nanowires have diameters between 57 and 101~nm.

% FIG
\begin{figure}[t]
\includegraphics[width=\columnwidth]{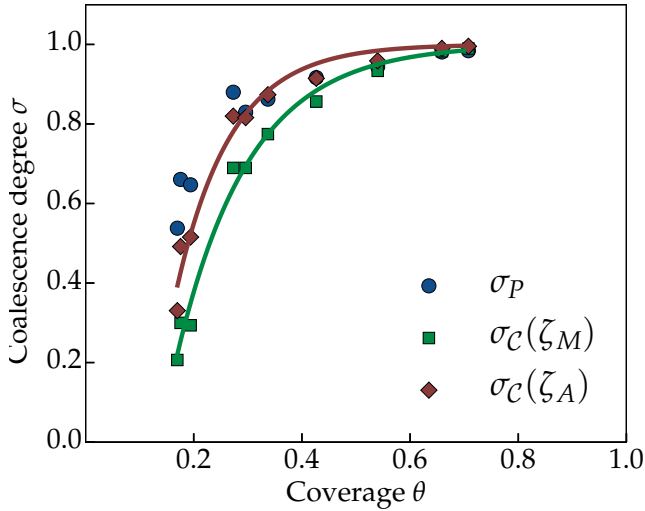} 
\caption{Coalescence degree $\sigma$ of the GaN nanowire ensembles under investigation as a function of the area coverage $\theta$. The symbols denote the results obtained by the criteria established in Fig.~\ref{fig5} for the sample series fabricated with systematically increased Ga/N flux ratio supplied during nanowire nucleation and growth. The lines are a guide to the eye.} 
\label{fig7} 
\end{figure}

Finally, we compare the criteria introduced above for nanowire ensembles with different fill factors or area coverages $\theta = A_T/A_{M}$ with the area $A_{M}$ covered by the entire sample. For this purpose, we investigate a series of samples grown under identical conditions except for the ratio of the impinging Ga and N fluxes, which was systematically varied between 0.2 and 1.4. Figure \ref{fig7} shows the coalescence degree of these samples versus their coverage. We have used both the expressions given above in Eqs.~(\ref{circularity_area}) and (\ref{equilaterality}), and we compare in addition two different circularity thresholds [specifically, $\zeta_{A} = 0.762$ corresponding to Fig.~\ref{fig3}(d) and $\zeta_{M} = 0.653$ corresponding to Fig.~\ref{fig3}(e)] to address the uncertainty related to the definition of the critical circularity separating single nanowires and coalesced aggregates as discussed above. The values obtained for $\zeta_{M}$ are naturally lower than those for a circularity of $\zeta_{A}$, but they still show the same trend: an unexpectedly steep increase of the coalescence degree with coverage, resulting in a coalescence degree of 0.8 for a coverage of 0.3. Note that this behavior is not a universal one but depends on growth conditions. For example, sample \textsf{A} exhibits a coalescence degree of only 0.75 for a coverage of 0.4.  

For a coverage exceeding 0.35, $\sigma_{\mathcal{C}}(\zeta_{A})$ is essentially identical to $\sigma_{P}$ which is determined without any element of ambiguity. For the lowest coverages, $\sigma_{P}$ is systematically higher than $\sigma_{\mathcal{C}}(\zeta_{A})$. In fact, the area-perimeter plot overestimates $\sigma_{P}$ for low degrees of coalescence for a simple mathematical reason. As evident from Fig.~\ref{fig5}(d)--\ref{fig5}(f), the data points exhibit a significant scatter which is caused by the coexistence of nanowires with identical area, but different shapes and thus perimeters. This coexistence is a direct consequence of the various distorted hexagonal shapes observed in the ensembles [cf.~Figs.~\ref{fig1}(a) and \ref{fig2}]. If the number of data points in the linear portion of the area-perimeter plot decreases, the least-square fit will tend to compensate for this decrease by averaging over the data points close to the critical perimeter $\eta$ separating the quadratic and linear portions of the plot. As a result, the fit will underestimate $\eta$ and overestimate $\beta$.

The latter quantity was found to have the same value for all nanowire ensembles with a coalescence degree larger than 0.6, namely, $1.0655 \pm 0.0005$. This finding reveals that the coalesced aggregates in GaN nanowire ensembles on Si(111) have the same fractal dimension regardless of the actual growth conditions.\cite{Voss_prl_1982} It furthermore offers the possibility to at least partly correct the deviation of $\sigma_{\mathcal{C}}(\zeta_{A})$ and $\sigma_{P}$ for low degrees of coalescence. If we set $\beta$ to the value given above, we indeed obtain higher values for $\eta$ and thus lower values for $\sigma_{P}$ compared to the case when $\eta$ is treated as free parameter. Even in this case, however, $\sigma_{P}$ still tends to be larger than $\sigma_{\mathcal{C}}(\zeta_{A})$. 

Considering these facts, we believe it to be advantageous to combine the two methods introduced in our work to determine the coalescence degree of nanowires. In fact, the two methods are to a certain degree complementary: the lower the value of $\sigma$, the higher is the accuracy of the circularity criterion. For large values of $\sigma$, on the other hand, the area-perimeter plot may be preferable since it does not require the definition of a threshold value which inevitably introduces an element of ambiguity. Since either of the two methods requires only the determination of the area and the perimeter of the nanostructures under investigation, it is actually straightforward to implement them in parallel.      

\section*{Conclusions}

Having established a method to characterize the coalescence degree of a GaN nanowire ensemble, we can begin to investigate several interesting topics in a systematic fashion.

For example, coalescence has often been ascribed to be caused by the tilt of neighboring nanowires.\cite{Consonni_apl_2009,Kaganer_prb_2012} However, a random tilt cannot possibly account for the high degree of coalescence obtained in the present work for simple geometrical reasons (a 1~$\mu$m long nanowire with a tilt of 1\grad bridges a distance of only 17~nm). In fact, in a recent work of Grossklaus \emph{et al.}\cite{Grossklaus_jcg_2013} also perfectly aligned nanowires were observed to coalesce, a process which was suggested to be driven by the reduction of total energy associated to the removal of free surfaces. This mechanism is actually well-known and has proven to be essential for the understanding of tensile stresses in polycrystalline materials.\cite{Nix_jmr_1999} Data such as shown in Fig.~\ref{fig7} will be useful to decide which mechanism is actually dominating the coalescence of GaN nanowires.

In this context, it would be highly interesting to study nanowires with a coverage below 0.2, a range left unexplored in the present paper. If the nucleation of nanowires is spatially uniform and random, their degree of coalescence should be finite for any coverage above zero. If nanowire nuclei avoid each other, however, a critical coverage is expected to exist below which no coalescence occurs. By studying the coalescence degree as a function of coverage, we hence gain information not only about the mechanisms governing nanowire coalescence, but also about those controlling nanowire nucleation.

Another important subject which may now be studied in more detail than previously possible is the impact of coalescence on the physical properties of the nanowires. The inhomogeneous strain detected in the x-ray diffraction experiments carried out on in Refs.~\citenum{Jenichen_nt_2011} and \citenum{Kaganer_prb_2012} has been suspected to arise at least partly from coalescence. In investigations of nanowire ensembles with low to medium coalescence degree,\cite{Jenichen_nt_2011} the net (homogenous) strain was found to be zero as expected,\cite{Brandt_prb_2010} but if nanowire coalescence is indeed driven by the minimization of free surface area, we would expect to see a tensile net strain developing for high coalescence degrees. In fact, such a strain state has recently been measured in GaN layers produced by the intentional coalescence of top-down produced GaN nanopillars.\cite{Hugues_jap_2013} Both the homogeneous and inhomogeneous strain should manifest themselves not only by a shift and broadening, respectively, of x-ray reflections, but also of phonon modes detected by Raman scattering experiments and excitonic transitions observed by photoluminescence spectroscopy.

Last but not least, coalescence may introduce not only strain, but also structural defects as most clearly documented by Grossklaus \emph{et al.}\cite{Grossklaus_jcg_2013} The threading dislocations in heteroepitaxial GaN films are known to induce nonradiative recombination, which may also prove true for the boundary dislocations at the coalescence boundaries of nanowires. Time-resolved photoluminescence spectroscopy performed for nanowire ensembles of different degree of coalescence could be helpful in establishing the actual impact of these dislocations. Since dislocations will only be formed if the tilt cannot be accommodated elastically, it is also of interest to compare samples with nanowires of different mutual misorientation but similar degree of coalescence. Well-aligned nanowires can be obtained, for example, by their epitaxial growth on coherent AlN layers on SiC(0001),\cite{Garrido_nl_2012} while nanowires formed on Si(001)\cite{Cerutti_apl_2006}, amorphous SiO$_{x}$\cite{Stoica_small_2008} or amorphous AlO$_{x}$\cite{Sobanska_jcg_2014} exhibit a fiber texture with poorly defined or even entirely random in-plane orientation.  

\begin{acknowledgement}
We thank Anne-Kathrin Bluhm for several scanning electron micrographs used in this work. We are also indebted to Henning Riechert for continuous encouragement and support and to Lutz Geelhaar for numerous valuable discussions and a critical reading of the manuscript. Finally, we gratefully acknowledge financial support of this work by the Deutsche Forschungsgemeinschaft through SFB 951.
\end{acknowledgement}

%%%%%%%%%%%%%%%%%%%%%%%%%%%%%%%%%%%%%%%%%%%%%%%%%%%%%%%%%%%%%%%%%%%%%
\bibliography{bibliography}

\end{document}